\documentclass[letterpaper, 10 pt, conference]{ieeeconf}  

\IEEEoverridecommandlockouts                              

\usepackage{graphics}       
\usepackage{epsfig}         
\usepackage{times}          
\usepackage{amsmath}        
\usepackage{amssymb}        
\usepackage{color}
\usepackage{array}
\usepackage{accents}

\usepackage{stackengine}
\usepackage{url}

\title{\LARGE \bf
Optimal terminal sliding mode control for second-order motion
systems}

\author{Michael Ruderman
\thanks{This work has received funding from the EUs
H2020-MSCA-RISE research and innovation programme under grant agreement No 734832.}%
\thanks{M. Ruderman is with Faculty of Engineering and
Science, University of Agder, Norway. Email:
        {\tt\small michael.ruderman@uia.no}}%
}

\begin{document}

\maketitle \thispagestyle{empty} \pagestyle{empty}

\begin{abstract}
Terminal sliding mode (TSM) control algorithm and its non-singular
refinement have been elaborated for two decades and belong, since
then, to a broader class of the finite-time controllers, which are
known to be robust against the matched perturbations. While TSM
manifold allows for different forms of the sliding variable, which
are satisfying the $q/p$ power ratio of the measurable output
state, we demonstrate that $q/p=0.5$ is the optimal one for the
second-order Newton's motion dynamics with a bounded control
action. The paper analyzes the time-optimal sliding surface and,
based thereupon, claims the optimal TSM control for the
second-order motion systems. It is stressed that the optimal TSM
control is fully inline with the Fuller's problem of optimal
switching which minimizes the settling time, i.e. with
time-optimal control of an unperturbed double-integrator. Is is
also shown that for the given plant characteristics, i.e. the
overall inertia and control bound, there is no need for additional
control parameters. The single surface design parameter might (but
not necessarily need to) be used for driving system on the
boundary layer of the twisting mode, or for forcing it to the
robust terminal sliding mode. Additional insight is given into the
finite-time convergence of TSM and robustness against the bounded
perturbations. Numerical examples with different upper-bounded
perturbations are demonstrated.
\end{abstract}

\bstctlcite{references:BSTcontrol}

\section{INTRODUCTION}
\label{sec:1}

Terminal sliding mode control belongs to the class of finite time
controllers \cite{haimo1986finite}, see e.g.
\cite{moulay2006finite} for survey, for which a finite-time
convergence can be guaranteed. The finite-time convergence is here
valid for both, the \emph{reaching phase} during which the state
trajectories reach the predefined sliding surface, and the
\emph{sliding phase} of a convergence to the state equilibrium. In
other words, when the control task is formulated as a
zero-reference and non-zero initial value problem, the controlled
system attains the origin in the finite-time. Yet, for the
reaching phase one refers to a finite-time convergence of the
sliding variable to zero, while for the second (sliding) phase a
finite-time convergence of the output tracking error to zero is
meant. For more details on the reaching and sliding phases and
convergence of the sliding mode control systems we refer to the
seminal literature e.g.
\cite{utkin1992,perruquetti2002,shtessel2014}.

Based on \cite{zak1988,Venkataraman92} the so-called terminal
sliding mode control has been proposed in \cite{zhihong1994} and,
later, brought into a non-singular form in \cite{feng2002}. The
difference between both will be summarized in the preliminaries
provided in Section \ref{sec:2}. Remarkable is a fact that for the
second-order systems, which are in focus of this work, the
terminal sliding mode surface, correspondingly control, coincides
with the Fuller's problem of optimal relay switching
\cite{fuller1960relay}. The latter has also been brought into
context and analyzed in terms of designing high-order sliding mode
control algorithms in \cite{dinuzzo2009higher}. It should be
recalled that for the second-order Newtonian dynamics of a
relative motion, it is always possible to bring the system to an
equilibrium in finite time, by using a bounded control, i.e.
through inherently bounded actuation forces. The well-known
approach to this problem leads to the co-called bang-bang control
\cite{bryson1975}, which is the time-optimal solution of an
unperturbed double-integrator control problem. We will also
briefly sketch it in Section \ref{sec:3} for convenience of the
reader.

In the rest of the paper, we will show that an optimal terminal
sliding mode control can be designed for motion systems with the
bounded perturbations and control actions, based on the
time-optimal switching of double-integrator. The main results with
analysis are given in Section \ref{sec:4}. Section \ref{sec:5}
demonstrates three numerical examples of the control performance
for different-type matched perturbations. The main conclusions are
briefly drawn in Section \ref{sec:6}.

\section{PRELIMINARIES}
\label{sec:2}

The so-called terminal sliding mode control suggests the
first-order terminal sliding variable, cf. \cite{feng2002},
\begin{equation}\label{eq:1:1}
s = x_2 + \beta x_1^{q/p},
\end{equation}
where $\beta > 0$ is a design parameter and $p > q$ are the
positive odd constants \cite{feng2002}. Recall that the terminal
sliding mode controllers belong to the class of second-order
sliding modes with finite-time convergence, see e.g.
\cite{levant1993sliding}. For a terminal sliding mode, the control
of a dynamic system $\ddot{x}_1 = \dot{x}_2 = u$, in the most
simple case, is given by
\begin{equation}\label{eq:1:2}
u = -\alpha \, \mathrm{sign}(s),
\end{equation}
where $\alpha > 0$ is the control parameter. The requirement on
$p,q$ to be odd \cite{feng2002} can be relaxed, and a frequently
encountered terminal sliding surface is written as, cf.
\cite{levant1993sliding},
\begin{equation}\label{eq:1:3}
s = x_2 + \beta \sqrt{|x_1|} \mathrm{sign}(x_1).
\end{equation}
By taking the time derivative of the sliding surface one can show
that, cf. \cite{fridman2015continuous},
\begin{equation}\label{eq:1:4}
\dot{s} = -\alpha \, \mathrm{sign}(s) + \beta x_2 \bigl(2
\sqrt{|x_1|} \bigr)^{-1}.
\end{equation}
One can recognize that a singularity occurs if $x_2 \neq 0$ and
$x_1=0$. This situation does not occur in an ideal sliding mode,
i.e. $s=0$. However, the terminal sliding mode control with
surface \eqref{eq:1:1} cannot guarantee an always bounded control
signal before the system is on the sliding manifold $s=0$, i.e.
during the reaching phase which depends on the initial conditions.
Once on the surface, the sliding variable dynamics \eqref{eq:1:4}
reduces to
\begin{equation}\label{eq:1:5}
2 \dot{s} = -2 \alpha \, \mathrm{sign}(s) - \beta^2 \,
\mathrm{sign}(x_1).
\end{equation}
Thereupon, one can show that the existence condition $s\dot{s} <
0$ of an ideal sliding mode is fulfilled for $\beta^2 < 2 \alpha$.
For that case, the trajectories of the system reach the surface
\eqref{eq:1:3} and remain there for all times afterwards. This
control system behavior is often denoted as a \emph{terminal
mode}, cf. \cite{sanchez2014lyapunov,fridman2015continuous}. When
the design parameters are assigned as $\beta^2 > 2 \alpha$, the
trajectories of the system do not remain on the sliding surface,
while the control still switches at $s=0$, according to
\eqref{eq:1:2}, \eqref{eq:1:3}. That case the control system
trajectories proceed in the so-called \emph{twisting mode}, cf.
\cite{sanchez2014lyapunov,fridman2015continuous}.

For overcoming the singularity of the sliding surface
\eqref{eq:1:1}, the non-singular terminal sliding mode control
with
\begin{equation}\label{eq:1:6}
s = x_1 + \beta^{-1} x_2^{p/q},
\end{equation}
has been proposed in \cite{feng2002}, while $\beta$, $p$, and $q$
parameters are kept the same as before. This control works around
the surface singularity problem while, at the same time,
\eqref{eq:1:1} and \eqref{eq:1:6} describe one and the same
switching surface, seen from the geometrical interpretation, in
the $(x_1, x_2)$ coordinates.

\section{TIME OPTIMAL CONTROL OF SECOND-ORDER MOTION SYSTEMS}
\label{sec:3}

We consider a generalized\footnote{We mean here the generalized
coordinates $x$ and forces $u$ for not explicitly distinguishing
between the translational and rotational degrees of freedom
(DOFs). As consequence, $m$ is used for denoting equally both
quantities, inertial mass and moment of inertia respectively.}
second-order motion system
\begin{equation}\label{eq:1}
\ddot{x}(t) = m^{-1} u(t)
\end{equation}
of an inertial body $m$, which is first not affected by any
counteracting, correspondingly disturbing, forces. For the sake of
simplicity and without loss of generality we focus on the 1DOF
motion, so that the scalar state variables of motion are $x_1=x$
and $x_2 = \dot{x}$. The control input force, here and for rest of
the paper, is constrained as $u \in [-U, +U]$. Here it is worth
emphasizing that the bounded (often denoted as \emph{saturated})
control signal, correspondingly actuation force, is in focus of
our optimal terminal sliding mode control. Consequently, $U$ will
appear as a system parameter in the following analysis and control
synthesis.

Assuming an initial state $[x_1(0), x_2(0)]^T$ and a final state
$[x_1(t_f), x_2(t_f)]^T$ the time-optimal control
\cite{fuller1960relay,bryson1975,geering2007optimal} will minimize
the cost criterion
\begin{equation}\label{eq:2}
J(u) = t_f = \int\limits_{0}^{t_f} dt,
\end{equation}
thus minimizing the overall convergence (or settling
\cite{fuller1960relay}) time $t_f$. Following the Pontryagin's
minimum principle and solving the constrained (by control limit
$U$ and initial and final conditions) optimization problem
\eqref{eq:2} leads to the Hamiltonian function
\begin{equation}\label{eq:3}
H = \lambda_0 + \lambda_1(t)x_2(t) + \lambda_2(t)u(t),
\end{equation}
with $\lambda_{0-2}$ to be the vector of Lagrange multipliers.
Minimizing the Hamiltonian function yields the time-optimal
control of the form, cf. with \cite{geering2007optimal},
\begin{equation}\label{eq:4}
u^o(t) = \left\{%
\begin{array}{ll}
    +U, & \hbox{for } \lambda_2^o(t) < 0, \\
    0, & \hbox{for } \lambda_2^o(t) = 0, \\
    -U, & \hbox{for } \lambda_2^o(t) > 0, \\
\end{array}%
\right.
\end{equation}
where $\lambda_2^o(t)$ is the control-related optimal Lagrange
multiplier. Substituting this control law into $x_2$-dynamics
results in a two-point boundary value problem, see
\cite{geering2007optimal} for details, from which the solution of
optimal Lagrange multiplier gives
\begin{equation}\label{eq:5}
\lambda_2^o(t) = -c_1^o \, t + c_2^o.
\end{equation}
Finally, the optimal constant values $(c_1^o, c_2^o) \neq (0,0)$
need to be found such that the two-point boundary value problem is
solved. With closer look on \eqref{eq:4}, \eqref{eq:5} one can
recognize that: (i) the motion system \eqref{eq:1} is always fully
either accelerated or decelerated, (ii) the calculated multiplier
\eqref{eq:5} determines the time instant of exactly one switching
between both control actions, e.g. from $+U$ to $-U$ when $x_1(0)
< x_1(t_f)$. Such control behavior is well known as the so-called
``bang-bang'' control, while the optimal switching time, resulting
from \eqref{eq:5}, is always depending on the initial and final
states. Here it is worth recalling that the time-optimal control
\eqref{eq:4}, solved in terms of a switching time instant
\eqref{eq:5}, was formulated as the Fuller's problem in
\cite{fuller1960relay} and further analyzed for an optimal sliding
mode control in \cite{dinuzzo2009higher}, in both cases in terms
of the state variables, correspondingly output tracking error.

For the controlled system \eqref{eq:1}, \eqref{eq:4} we are next
to analyze the decelerating trajectory, i.e. the state trajectory
after the time-optimal switching at $t=\tau$, and that for
$u(t)=-U \; \forall \; \tau \leq t \leq t_f$. For the sake of
simplicity and without loss of generality we will assume
$x_1(t_f)=x_2(t_f)=0$, that for the rest of the paper, meaning
zero equilibrium point. Consequently, all our following analysis
and developments reduce to the problem of convergence to a stable
equilibrium in the origin, while other control tasks with
$\bigl(x_1(t_f),x_2(t_f)\bigr) \neq 0$ can be converted to that
one by an appropriate transformation of the state coordinates. For
the phase-plane analysis, the system \eqref{eq:1} with $u=-U$ can
be rewritten as
$$
\dot{x}\,d \dot{x} = -U m^{-1} dx,
$$
and after integration of both sides and substitution of $x_1$ and
$x_2$ one obtains the parabolic trajectory
\begin{equation}\label{eq:6}
x_1 = -0.5m \, U^{-1} x_2^2.
\end{equation}
\begin{figure}[!h]
\centering
\includegraphics[width=0.7\columnwidth]{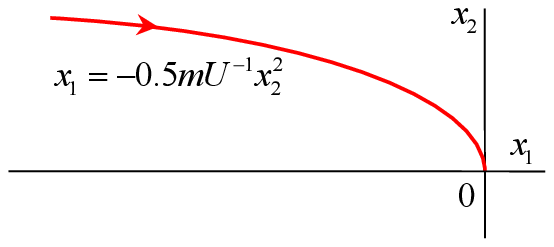}
\caption{Decelerating trajectory of time optimal control
\eqref{eq:4}.} \label{fig:1}
\end{figure}
Obviously, the upper branch of the $(x_1,x_2)$-parabola
constitutes the motion trajectory until $t=t_f$. The motion
trajectory is steered with the maximal possible deceleration
forced by $u(t)=-U$. An example of the converging state
trajectory, driven by the time-optimal control \eqref{eq:4} during
the decelerating phase, is shown in Fig. \ref{fig:1}.

\section{OPTIMAL TERMINAL SLIDING MODE CONTROL}
\label{sec:4}

Now, with the preliminaries from Section \ref{sec:1} and time
optimal control summarized in Section \ref{sec:2}, we are in the
position to formulate the optimal terminal sliding mode control
for the perturbed second-order motion systems.

The proposed optimal terminal sliding surface is given by
\begin{equation}\label{eq:7}
s = x_1 + \alpha m  U^{-1} x_2^2 \, \mathrm{sign}(x_2),
\end{equation}
with $\alpha > 0$ to be the single (optional) design parameter.
Note that the proposed switching, correspondingly sliding, surface
\eqref{eq:7}, cf. \cite[eq.~(53)]{fuller1960relay}, has the same
form as the non-singular terminal sliding mode algorithm, see
\cite{fridman2015continuous}, with $q/p=0.5$ rate cf. Section
\ref{sec:2}. Important to notice is that both surface shaping
factors, $U$ and $m$, do not represent any free parameters to be
tuned. They are rather the inherent physical quantities which are
characterizing the motion system plant \eqref{eq:1}. One can
recognize that the trajectory \eqref{eq:6} (of a time-optimal
controlled system) coincides with the sliding surface \eqref{eq:7}
for $\alpha = 0.5$. Further we will show that for that case, the
boundary layer of the so-called twisting mode appears, cf.
\cite{sanchez2014lyapunov,fridman2015continuous}. The control
value is
\begin{equation}\label{eq:8}
u(s) = -U \mathrm{sign}(s),
\end{equation}
in which the control gain factor is the maximal possible one, i.e.
inherently limited by the constrained actuator force.

The sufficient condition for existence of the terminal sliding
mode on the surface \eqref{eq:7} can be shown by assuming the
candidate Lyapunov function
\begin{equation}\label{eq:9}
V = 0.5 s^2,
\end{equation}
for which
\begin{equation}\label{eq:10}
0.5 \frac{d}{dt} s^2 < 0
\end{equation}
is, respectively, required. That means for the trajectory which
has once reached and crossed the surface $s=0$ at $t=t_r$, it
remains on it $\forall \: t > t_r$. The corresponding motion along
$s=0$, denoted as terminal mode, requires $s\dot{s}<0$ to be
always fulfilled. For showing it, we exemplary consider $x_2 > 0$,
thus focusing, yet without loss of generality, on the upper
parabolic branch in the second quadrant of the phase-plane only.
Next we summarize $\alpha m U^{-1} = : C$, for the sake of
simplicity, and with these simplifications in mind the terminal
sliding surface \eqref{eq:7} can be written as a parabolic
equation $s=x_1 + C x_2^2$. Taking the time derivative and
substituting the control \eqref{eq:8} into the dynamics
\eqref{eq:1} one obtains
\begin{equation}\label{eq:11}
\dot{s} = x_2 - 2 m^{-1} C U x_2 \, \mathrm{sign}(s).
\end{equation}
Multiplying both sides of \eqref{eq:11} with $s$, the existence
condition \eqref{eq:10} results in
\begin{equation}\label{eq:12}
x_2 s - 2 m^{-1} C U x_2 |s| < 0.
\end{equation}
Substituting back the plant parameters, instead of $C$, and
solving the inequality \eqref{eq:12} with respect to the single
free factor $\alpha$, the existence condition of the optimal
terminal sliding mode results in
\begin{equation}\label{eq:12x}
\alpha > 0.5.
\end{equation}

Important to notice is that if the design parameter is selected as
$0 < \alpha \leq 0.5$, the existence condition \eqref{eq:10} of
the sliding mode becomes violated, and the trajectories do not
remain on $s=0$ upon crossing. Nevertheless, the manifold
\eqref{eq:7} appears further on as a switching surface, and the
trajectories will reach asymptotically the origin according to
\eqref{eq:1} and \eqref{eq:8}. This convergence behavior appears
as a twisting mode, in which the frequency of switching (upon
crossing the surface \eqref{eq:7} with $0 < \alpha \leq 0.5$)
increases towards infinity as the trajectory approaches the
origin. Recall that the trajectory converges towards origin while
circulating around it. The convergence of the twisting mode can be
directly proved geometrically, since $\|x_1,x_2 \|(t_{s2}) <
\|x_1,x_2 \|(t_{s1})$ for all pairs of two consecutive switching
(correspondingly $s$-crossing) at $t_{s2} > t_{s1}$. This is valid
as long as $\alpha > 0$. When $\alpha \rightarrow 0$, the
switching surface $s \rightarrow x_1$ and the twisting occurs
without convergence, while representing the boundary case of a
harmonic oscillator. The latter is nothing but the
double-integrator with a zero-crossing relay in feedback. The
stability of the twisting mode has also been explicitly addressed
in \cite{sanchez2014lyapunov} by means of the Lyapunov's method.
It is also worth noting that for $0.25 \leq \alpha \leq 0.5$, the
sliding surface still falls into class of the Fuller's problem,
cf. \cite{fuller1960relay,dinuzzo2009higher}. Also one can easily
recognize that the positive $\alpha$ appears, from a plant
parameters viewpoint, as a gain factor which captures the
multiplicative uncertainties in $m$ and $U$, cf. \eqref{eq:7}.

\begin{figure}[!h]
\centering
\includegraphics[width=0.98\columnwidth]{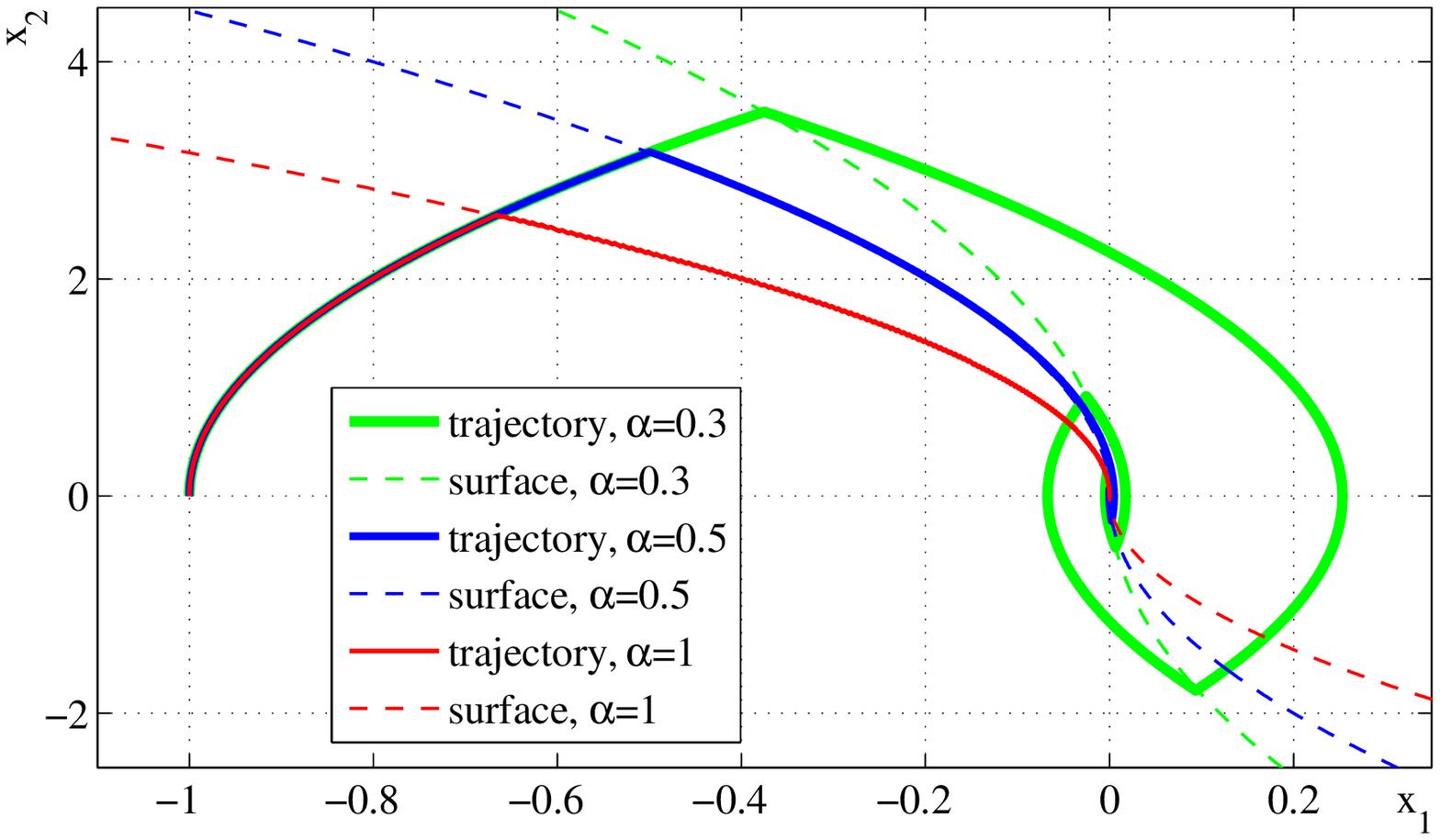}
\caption{Sliding surface and trajectories for different $\alpha$
values.} \label{fig:2}
\end{figure}

The phase-portrait with trajectories of the system \eqref{eq:1},
\eqref{eq:8} is exemplary shown in Fig. \ref{fig:2} together with
the corresponding surface \eqref{eq:7}, that for the different
$\alpha$-values.
Here the system parameters are assigned as in the numerical
examples provided in Section \ref{sec:5}, while the initial values
$\bigl(x_1(t_0),x_2(t_0)\bigr)=(-1,0)$ are assumed for all three
$\alpha$ values. One can directly notice that once the switching
surface is reached, the control with $\alpha = 0.3$, which
violates \eqref{eq:12x}, does not stay sliding on the surface, and
the trajectory converges in a twisting mode. Note that here the
chattering, due to non-ideal switching and numerical computational
issues, does not appear until the states reach the origin, meaning
until $\bigl(x_1(t_f),x_2(t_f)\bigr) \rightarrow \mathbf{0}$.

In order to establish the reachability condition, see
\cite{edwards1998,shtessel2014}, one consider the candidate
Lyapunov function \eqref{eq:9} for which the following conditions
should be satisfied: (i) $V(s)$ is positive definite $\forall \,
s$, (ii) $\dot{V} < 0$ for $s \neq 0$. For showing the finite-time
convergence and, therefore, global stability the so-called
$\eta$-reachability \cite{edwards1998} condition
\begin{equation}\label{eq:13}
s\dot{s} < -\eta |s|
\end{equation}
can be evaluated, in which $\eta > 0$ is a small constant which
ensures the reachability condition is satisfied. One slightly
modifies the above condition (ii) to be
\begin{equation}\label{eq:14}
\dot{V} \leq - \gamma \sqrt{V}, \quad \gamma = \sqrt{2} \eta.
\end{equation}
Integrating both sides of inequality \eqref{eq:14} one obtains,
cf. \cite{shtessel2014},
\begin{equation}\label{eq:15}
\sqrt{V}(t) \leq -0.5 \gamma t + \sqrt{V(0)}.
\end{equation}
Since reaching the sliding surface at finite-time $t_r$ means
$V(t_r) = 0$, one can obtain, out from \eqref{eq:15}, the
following
\begin{equation}\label{eq:16}
t_r \leq \gamma^{-1} 2 \sqrt{V(0)} = \gamma^{-1} 2 \sqrt{0.5}
|s(0)|.
\end{equation}
One can recognize that a bounded finite time can be guaranteed by
\eqref{eq:16} and it depends on the initial value of the sliding
manifold only. In order to evaluate the $\eta$-reachability
condition for the control system \eqref{eq:1}, \eqref{eq:7},
\eqref{eq:8}, we first rewrite \eqref{eq:13} as
\begin{equation}\label{eq:17}
\mathrm{sign}(s)\dot{s} < -\eta.
\end{equation}
Taking the time derivative of \eqref{eq:7} and substituting the
control system dynamics \eqref{eq:1}, \eqref{eq:8}, one obtains
\begin{equation}\label{eq:18}
\dot{s} = x_2 - 2\alpha |x_2| \mathrm{sign}(s).
\end{equation}
Combining \eqref{eq:17} and \eqref{eq:18} and separating the
variables we achieve the $\eta$-reachability condition as
\begin{equation}\label{eq:19}
x_2 \mathrm{sign}(s) < -\eta + 2\alpha |x_2|.
\end{equation}
One can see that it is always possible to find a small positive
constant $\eta$ such that the condition \eqref{eq:19} is fulfilled
$\forall \: x_2$.

Once the existence and reachability conditions, \eqref{eq:12x} and
\eqref{eq:19} respectively, are satisfied it is indicative to
analyze the control system robustness against the perturbations.
The external and internal perturbations\footnote{Here we refer to
the internal perturbations as well, since these may arise from the
plant uncertainties and internal dynamics which are not captured
by the motion system in form of the double-integrator
\eqref{eq:1}.} matched by the control will directly affect the
right-hand side of \eqref{eq:1} and, in worst case, produce a
counteracting (disturbing) force during the reaching phase, for
which $|u|(t) = U$ during the time $0 < t < t_r$. On the other
hand, the sudden `fast' perturbations, once they are occurring
during the sliding mode, will diverge the trajectories from the
sliding manifold, thus giving rise (again) to the reachability
problem. Such transient loss of the sliding mode leads, in turn,
to $u(t) = \mathrm{const} \in \{-U, U\}$ during the time $t_d < t
< t_r$, where $t_d > 0$ is the last instant of losing the sliding
manifold $s$ due to appearance of the perturbations. In order to
guarantee that the optimal terminal sliding mode control can
compensate for the unknown perturbations $\xi(t)$, the single
boundedness condition
\begin{equation}\label{eq:20}
|\xi| < U,
\end{equation}
is required for all times $t > 0$ of the control operation.

In the following, we will demonstrate different examples of the
optimal terminal sliding mode control applied to the perturbed
motion system. For that purpose, the system plant is extended by
the matched upper bounded perturbation \eqref{eq:20}, thus
resulting in the modified \eqref{eq:1} which becomes
\begin{equation}\label{eq:21}
\ddot{x}(t) = m^{-1} \bigl(u(t) + \xi(t)\bigr).
\end{equation}

\section{NUMERICAL EXAMPLES}
\label{sec:5}

The assumed numerical parameters, here and for the rest of the
paper, are $m=0.1$, $U=1$, $\alpha=0.6$. All the controlled motion
trajectories start at the initial state
$\bigl(x_1(t_0),x_2(t_0)\bigr)=(-1,0)$, while the control set
point is placed in the origin, meaning
$\bigl(x_1(t_f),x_2(t_f)\bigr)_{ref}=\mathbf{0}$. The implemented
and executed simulations are with 1kHz sampling rate and the (most
simple) first-order Euler numerical solver.

\subsection{Motion system with nonlinear friction}
\label{sec:5:1}

The motion system dynamics \eqref{eq:1} is first considered as
being affected by the nonlinear Coulomb friction $f(x_2,t)$ which,
moreover, includes the continuous presliding transitions, see
\cite{ruderman2017} for details. It is worth emphasizing that the
modeled Coulomb friction force is not discontinuous at zero
velocity crossing, while a saturated friction force at
steady-state results in $F_c\, \mathrm{sign}(\dot{x})$, where $F_c
> 0$ is the Coulomb friction constant. The friction force, which
appears as an internal perturbation $\xi=-f$, is assumed to be
bounded by the half of the control amplitude, i.e. $F_c=0.5$.
\begin{figure}[!h]
\centering
\includegraphics[width=0.98\columnwidth]{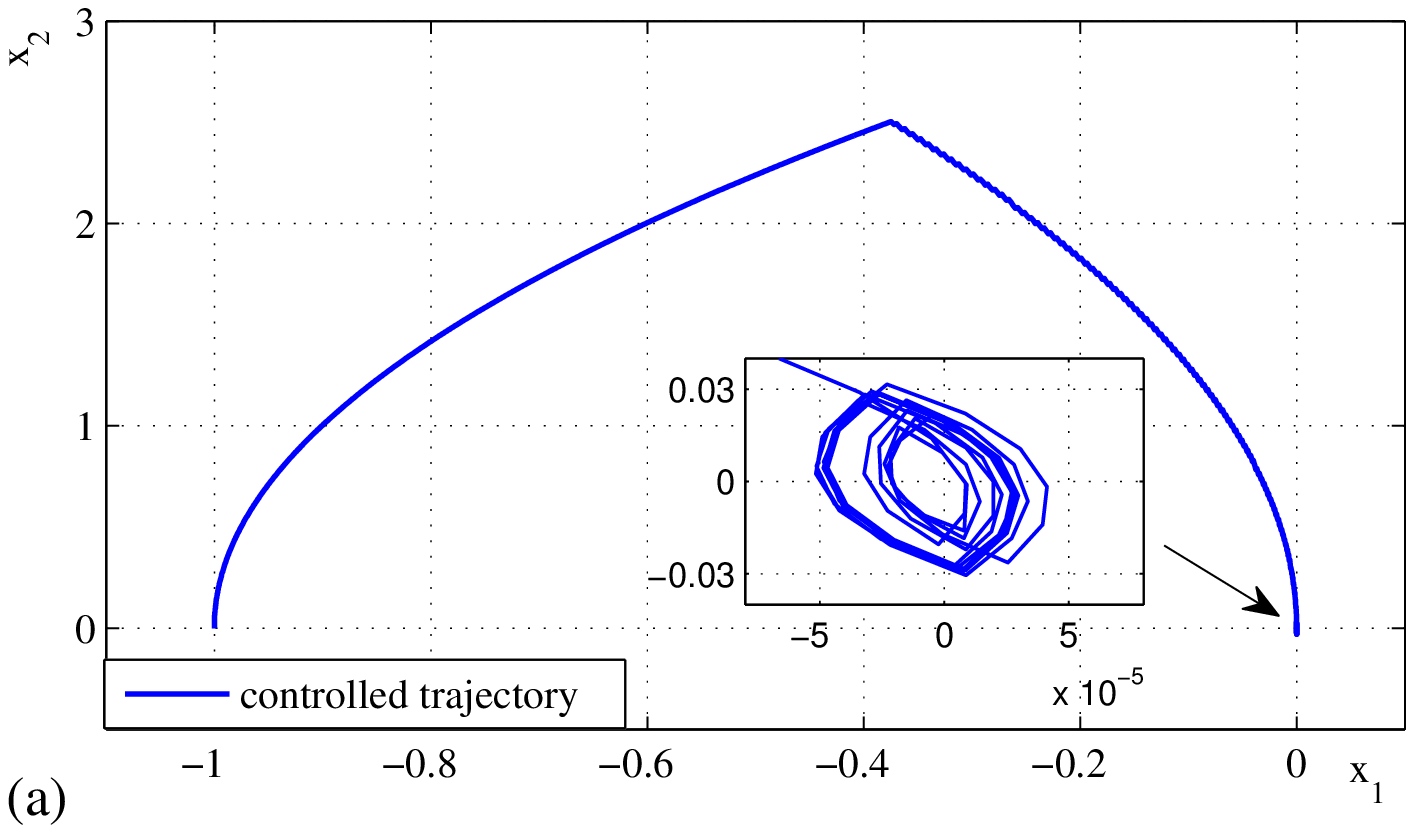}
\includegraphics[width=0.98\columnwidth]{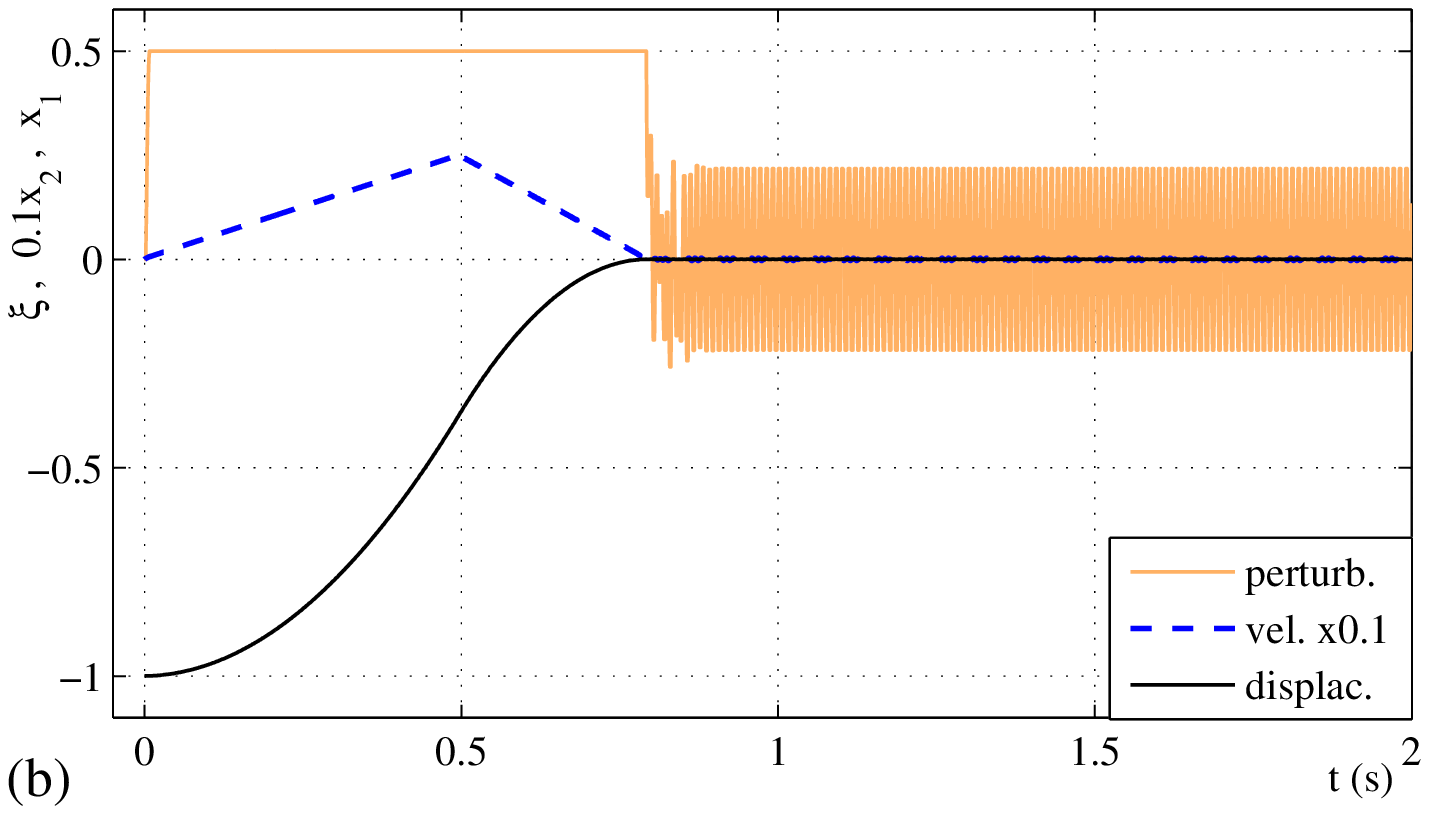}
\caption{Phase portrait of state trajectory (a) and convergence of
output tracking variable $x_1(t)$ and down-scaled $x_2(t)$-state
over perturbation (b).} \label{fig:3}
\end{figure}
The controlled state trajectory is shown in Fig. \ref{fig:3} (a),
while the region around origin is additionally zoomed-in for the
sake of visualization. One can see that after the state
convergence, in the sliding mode, a stable low-amplitude
($\times10^{-5}$) limit cycle around zero appears due to the
by-effects caused by the nonlinear presliding \cite{ruderman2017}
friction. This is not surprising since $F_c = 0.5 U$, and no
explicit presliding friction compensation is performed by the
terminal sliding mode control which is only switching between $u =
\pm U$. The convergence of both state variables ($x_2$ is
down-scaled by factor 0.1 for the sake of visualization) is shown
in Fig. \ref{fig:3} (b) over the progress of frictional
perturbation. It can also be side-noted that the feasibility of
the Coulomb friction compensation by means of a classical
(first-order) sliding mode control with discontinuous control
action was experimentally demonstrated, in combination with a
time-optimal bang-bang strategy, in a case study presented in
\cite{ruderman2014}.

\subsection{Motion system with harmonic perturbation}
\label{sec:5:2}

\begin{figure}[!h]
\centering
\includegraphics[width=0.98\columnwidth]{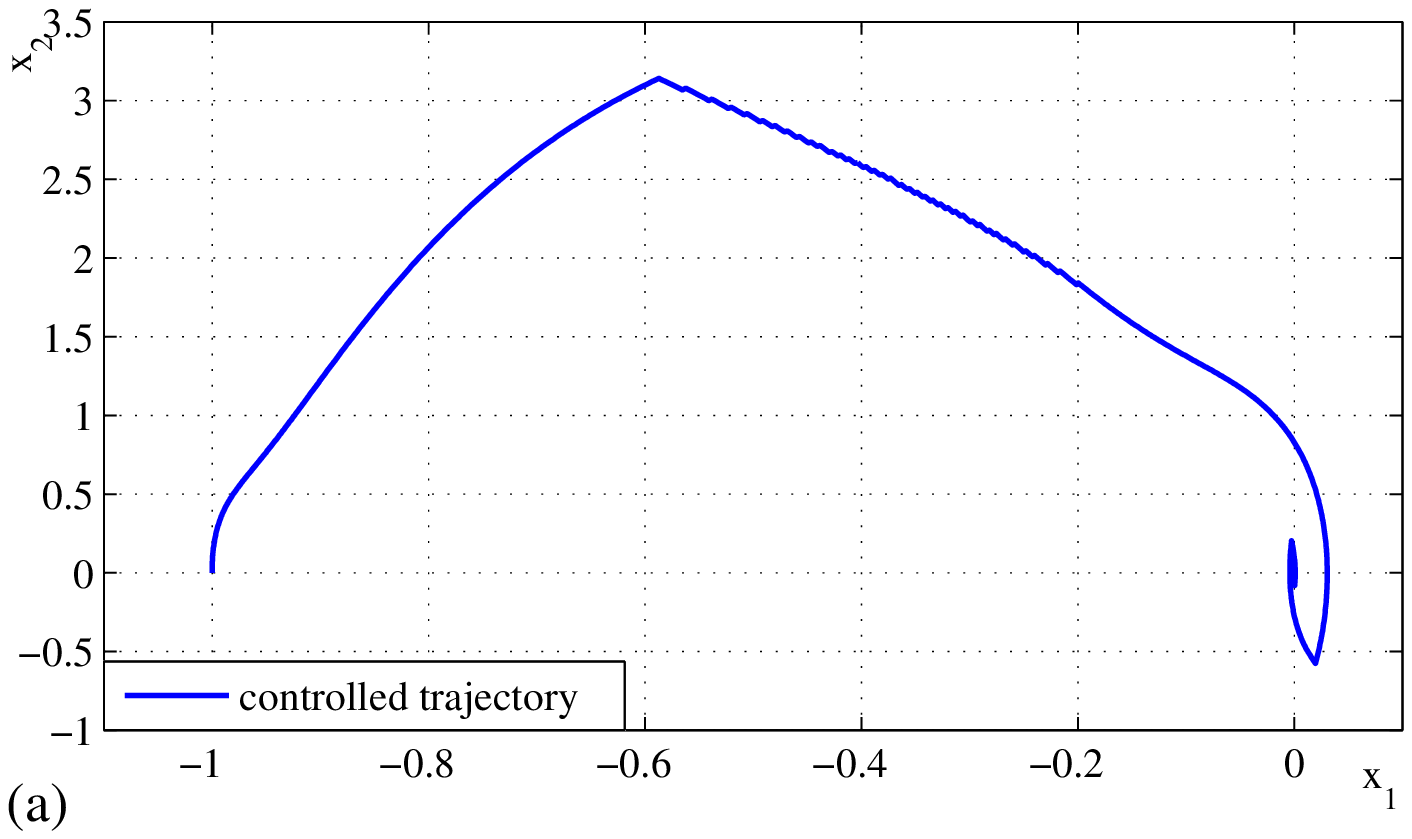}
\includegraphics[width=0.98\columnwidth]{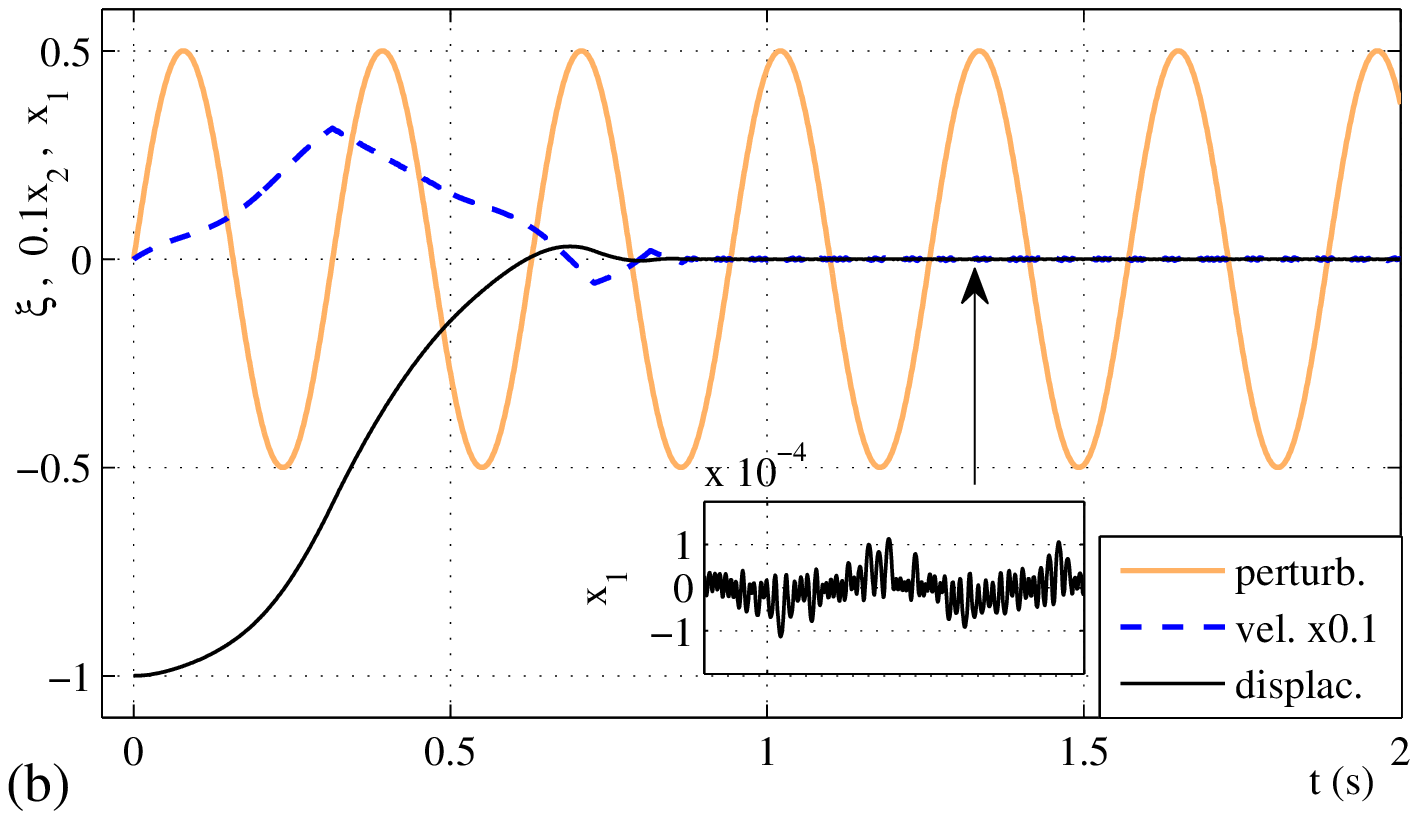}
\caption{Phase portrait of state trajectory (a) and convergence of
output tracking variable $x_1(t)$ and down-scaled $x_2(t)$-state
over perturbation (b).} \label{fig:4}
\end{figure}
As next, we assume an internal (or external) perturbation to be a
periodic function of time, modeled by $\xi(t)= 0.5 \sin(20 t)$.
Note that such disturbing harmonic forces (or torques) may occur
in various types of the motors, or more generally actuators, for
example see \cite{RuderEtAl13}. The sinusoidal amplitude is
selected to be the half of the control value, and the angular
frequency of 20 rad/s is assigned for providing a sufficient
number of periods during both the transient and steady-state
response. The controlled state trajectory and the converged state
and perturbation values are shown in Fig. \ref{fig:4}. Also the
zoomed-in steady-state output error is plotted.

\subsection{Motion system with random binary perturbation}
\label{sec:5:3}

Finally, we consider the motion system affected by an external
stepwise perturbation which is appearing in form of a random
binary signal. The perturbation amplitude is kept at the same
level as in both previous examples (cf. Sections \ref{sec:5:1} and
\ref{sec:5:2}), so that the perturbed plant \eqref{eq:21} has
$\xi(t) \in \{-0.5, 0.5\}$ which is a random realization of the
binary variable.
\begin{figure}[!h]
\centering
\includegraphics[width=0.98\columnwidth]{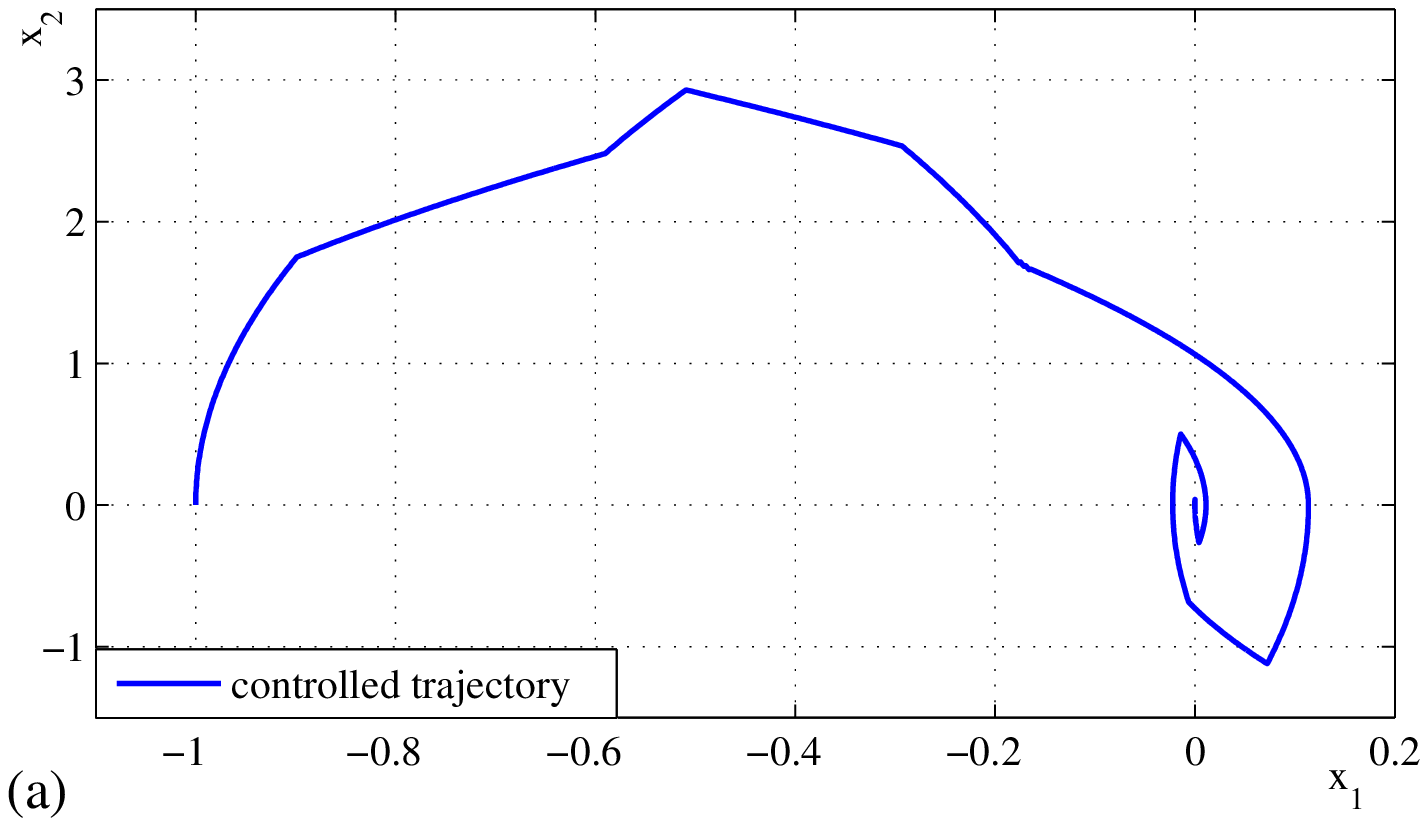}
\includegraphics[width=0.98\columnwidth]{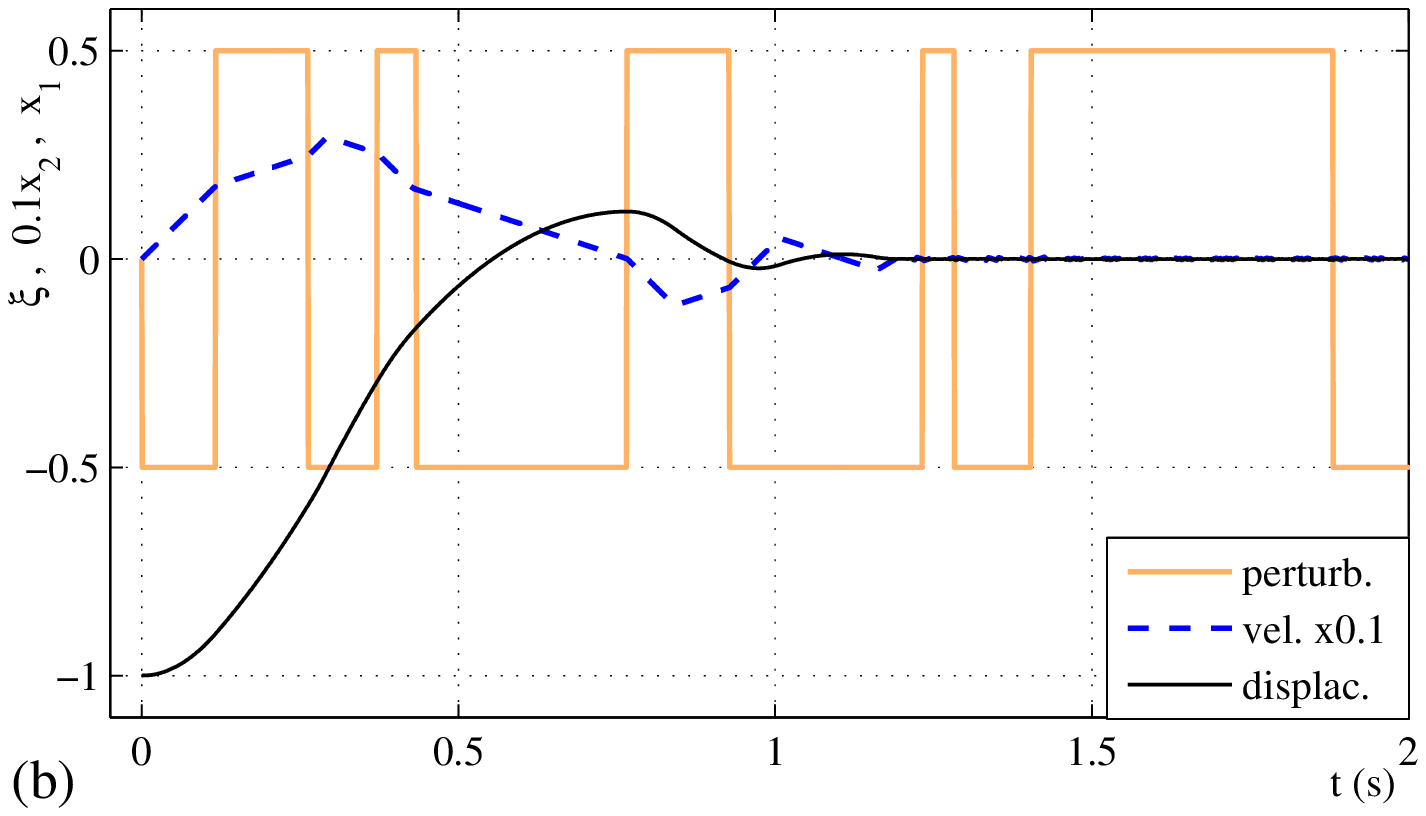}
\caption{Phase portrait of state trajectory (a) and convergence of
output tracking variable $x_1(t)$ and down-scaled $x_2(t)$-state
over perturbation (b).} \label{fig:5}
\end{figure}
The controlled state trajectory is shown in Fig. \ref{fig:5} (a),
and, alike, in (b) the states convergence and the perturbation
value are plotted together as the time series. One can recognize
that the motion trajectory is highly perturbed. Yet, both states
converge to zero, piecewise in the sliding mode and piecewise in
the twisting-like mode. The latter appears transiently each time
the binary perturbation changes the sign. It can be stressed that
from all three types of the perturbation examples, this one
constitutes the `worst' case which is largely affecting the
control system. This is not surprising since the differential
changes of the $\xi$-amplitude are of the same magnitude as $U$.
The stepwise perturbation $\xi(t)$ does not allow system to
continuously staying in the sliding mode and forces it multiple
times into the reaching phase. This is directly reflected in the
shape of $(x_1,x_2)$ trajectory, see Fig. \ref{fig:5} (a). At the
same time, one can recognize that the states remain in vicinity to
the origin, once reaching it and despite the further stepwise
changes of $\xi(t)$, see Fig. \ref{fig:5} (b) for the times $t >
1.25$ sec.

\section{CONCLUDING REMARKS}
\label{sec:6}

This paper described and analyzed the optimal terminal sliding
mode control aimed for the second-order motion systems with
matched and bounded perturbations. The control scheme uses the
time-optimal trajectory of the double-integrator dynamics, for
which only one dedicated switching provides the maximal possible
acceleration and deceleration in presence of the inherent control
bounds. The proposed optimal terminal sliding mode has the single
free (and rather optional) design parameter $\alpha$, which allows
for both, the terminal and twisting control modes. The
time-optimal twisting mode on the boundary layer, i.e. for
$\alpha=0.5$, is in particular interesting for those motion
control applications where a frequent sliding mode switching of
the control signal may be undesirable, like in the hydraulic
control valves \cite{ruderman2019}. Here we however abstain from
discussions on the pros and cons of the discontinuous and
continuous sliding mode controllers, see \cite{perez2019} for
details. We focus rather on the analysis and prove of existence
and reachability conditions of the proposed optimal terminal
sliding mode control. The shown developments and the determined
optimal sliding surface are fully inline with the Fuller's
problem. It was shown that for a motion dynamics with the given
inertial and actuator saturation parameters, there is no other
time-optimal surface than \eqref{eq:7}. All $\alpha > 0$ provide a
stable state convergence to the origin, while the $\alpha$-values
around boundary layer (i.e. $\alpha=0.5$) are particularly
interesting for the control application. These allow finding a
trade-off between a time-optimal convergence with minimal
switching and ability to capture multiplicative uncertainties of
the inertial and saturation parameters. Three illustrative
examples of the bounded internal and external perturbations
demonstrated the control efficiency in a numerical setup.

\bibliographystyle{IEEEtran}
\bibliography{references}

\end{document}